\documentclass[a4paper]{jpconf}
\usepackage{graphics}
\usepackage{enumerate}
\usepackage{amssymb}
\usepackage{epsfig,euscript,array}
\usepackage{axodraw}
\usepackage{subfigure}
\usepackage{color}
\usepackage{graphicx}
\begin{document}
\setlength{\footskip}{1cm}

\vspace*{-2.0cm}\noindent\hspace*{13cm}DESY 06-087\\
\hspace*{13cm}IPPP/06/34\\
\hspace*{13cm}DCPT/06/68
\vspace*{-0.7cm}

\title{Photon Defects in Noncommutative Standard Model Candidates\footnote{This is based on a talk given by J. Jaeckel at the CORFU 2005 Satellite workshop: ``Noncommutative Geometry in Field and String Theories''. For additional details see \cite{Jaeckel:2005wt,AJKR,AJKR2}.}}

\author{S. A. Abel$^a$, J. Jaeckel$^b$, V. V.  Khoze$^a$ and A. Ringwald$^b$}
\address{$^a$Center for Particle Theory, Durham University, Durham, DH1 3LE, UK}
\address{$^b$Deutsches Elektronen-Synchrotron DESY, Notkestrasse 85, D-22607  Hamburg, Germany}

\ead{s.a.abel@durham.ac.uk, jjaeckel@mail.desy.de, valya.khoze@durham.ac.uk, andreas.ringwald@desy.de}

\begin{abstract}
Restrictions imposed by gauge invariance in noncommutative spaces
together with the effects of ultraviolet/infrared mixing lead to
strong constraints on possible candidates for a noncommutative
extension of the Standard Model. We study a general class of
noncommutative models consistent with these restrictions.
Specifically we consider models based upon a gauge theory with the
gauge group ${\rm U}(N_1)\times {\rm U}(N_2) \times \ldots \times
{\rm U}(N_m)$ coupled to matter fields transforming in the
(anti)-fundamental, bi-fundamental and adjoint representations. We
pay particular attention to overall trace-U(1) factors of the
gauge group which are affected by the ultraviolet/infrared mixing.
Typically, these trace-U(1) gauge fields do not decouple
sufficiently fast in the infrared, and lead to sizable Lorentz
symmetry violating effects in the low-energy effective theory. In
a 4-dimensional theory on a continuous space-time making these
effects unobservable would require making the effects of
noncommutativity tiny, $M_{\rm{NC}}\gg M_{\rm{P}}$. This severely
limits the phenomenological prospects of such models. However,
adding additional universal extra dimensions the trace-U(1)
factors decouple with a power law and the constraint on the
noncommutativity scale is weakened considerably. Finally, we
briefly mention some interesting properties of the photon that
could arise if the noncommutative theory is modified at a high
energy scale.
\end{abstract}

\section{Introduction}
\pagestyle{plain}
Gauge theories on spaces with noncommuting coordinates,
\begin{equation}
[x^\mu,x^\nu]=i\,\theta^{\mu\nu} \ ,
\end{equation}
provide a very interesting
new class of quantum field theories with intriguing and sometimes unexpected features.
These noncommutative models can arise naturally as low-energy effective theories from string
theory and D-branes. As field theories they must satisfy a number of restrictive
constraints detailed below, and this makes them particularly interesting and challenging
for purposes of particle physics model building.
For general reviews of noncommutative gauge theories the reader can consult e.g.
Refs.~\cite{Seiberg:1999vs,Douglas:2001ba,Szabo:2001kg}.

There are two distinct approaches used in the recent literature for constructing
quantum field theories on noncommutative spaces. The first approach uses the Weyl-Moyal
star-products to introduce noncommutativity. In this case,
noncommutative field theories are defined by replacing the ordinary
products of all fields in the Lagrangians of their commutative counterparts
by the star-products
\begin{equation}
(\phi * \varphi) (x) \equiv \phi(x)\  e^{{i\over 2}\theta^{\mu\nu}
\stackrel{\leftarrow}{\partial_\mu}
\stackrel{\rightarrow}{\partial_\nu}} \  \varphi(x) \ . \label{stardef}
\end{equation}
Noncommutative theories in the Weyl-Moyal formalism can be viewed as field theories on
 ordinary commutative spacetime. For example,  the noncommutative pure
gauge theory action is
\begin{equation}
S = -{1\over 2g^2}\int d^{4} x \ \Tr ( F_{\mu \nu}*  F^{\mu \nu}
 ) \ , \label{pureym}
\end{equation}
where
the commutator in the field strength also contains the star-product. The important feature
of this approach is the fact that phase factors in the star-products are not expanded in powers of $\theta$ and
the $\theta$ dependence in the Lagrangian is captured entirely. This ability to work to all orders in $\theta$
famously gives rise to the ultraviolet/infrared (UV/IR) mixing ~\cite{Minwalla:1999px,Matusis:2000jf}
in the noncommutative quantum field theory which we will review below.

The second indirect approach to noncommutativity does not employ star-products. It instead
relies \cite{Madore:2000en,Calmet:2001na}
 on the Seiberg-Witten map which represents noncommutative fields as
a function of $\theta$ and ordinary commutative fields.
This approach essentially reduces noncommutativity to an introduction
of an infinite set of higher-dimensional (irrelevant) operators, each suppressed
by the corresponding power of $\theta$, into the action.
There are two main differences compared to the Weyl-Moyal approach. First, in practice one always works with the first few terms in the power series in $\theta$ and in this setting the UV/IR mixing cannot be captured. Second,
the Seiberg-Witten map is a non-linear field transformation. Therefore, one expects a non-trivial Jacobian and possibly a quantum theory different from the one obtained in the Weyl-Moyal approach.
In this paper we will use the original direct formulation of the theory
on a noncommutative space in terms of the Weyl-Moyal star product.

In the context of Weyl-Moyal
noncommutative Standard
Model building, a number of features of noncommutative
gauge theories have to be taken into account which are believed
to be generic~\cite{Khoze:2004zc}:
\begin{enumerate}[1.\,]
\item{} the mixing of ultraviolet and infrared effects~\cite{Minwalla:1999px,Matusis:2000jf}
        and the asymptotic decoupling of U(1) degrees of freedom~\cite{Khoze:2000sy,Hollowood:2001ng} in the
        infrared;
\item{} the gauge groups are restricted to U($N$) groups~\cite{Matsubara:2000gr,Armoni:2000xr} or products of thereof;
\item{} fields can transform only in (anti-)fundamental, bi-fundamental
        and adjoint representations~\cite{Gracia-Bondia:2000pz,Terashima:2000xq,Chaichian:2001mu};
\item{} the charges of matter fields are restricted~\cite{Hayakawa:1999zf}
        to $0$ and $\pm 1$,
        thus requiring extra care in order to give fractional electric charges
        to the quarks.
\end{enumerate}

Building upon an earlier proposal by Chaichian {\it{et al.}} \cite{Chaichian:2001py},
the authors of Ref.~\cite{Khoze:2004zc}
constructed an example of a noncommutative
embedding of the Standard Model with the purpose to satisfy all the requirements listed above.
The model of \cite{Khoze:2004zc} is based on the gauge group ${\rm{U}}(4)\times {\rm{U}}(3) \times {\rm{U}}(2)$
with matter fields transforming
in noncommutatively allowed representations. Higgs fields break the noncommutative gauge group down to
a low-energy commutative gauge theory which includes
the Standard Model group ${\rm{SU}}(3)\times {\rm{SU}}(2) \times {\rm{U}}(1)_Y$.
The ${\rm{U}}(1)_Y$ group here corresponds to
ordinary QED, or more precisely to the hypercharge $Y$
Abelian gauge theory. The generator of ${\rm{U}}(1)_Y$ was constructed from a linear combination of
{\it traceless} diagonal generators of the microscopic theory ${\rm{U}}(4)\times {\rm{U}}(3) \times {\rm{U}}(2).$
Because of this, the UV/IR effects -- which can affect only the overall trace-${\rm{U}}(1)$ subgroup of
each ${{\rm{U}}} (N)$ -- were not contributing to the hypercharge ${\rm{U}}(1)_Y.$ However some of the
overall trace-${\rm{U}} (1)$ degrees of freedom
can survive the Higgs mechanism and thus contribute to the low-energy
effective theory, in addition to the Standard Model fields. These additional
trace-${\rm{U}}(1)$ gauge fields logarithmically decouple from the low-energy effective theory
and were neglected in the analysis of Ref.~\cite{Khoze:2004zc}.
Here, we take these effects into account.

We will find that
the noncommutative model building constraints,
and, specifically, the UV/IR mixing effects in the trace-U(1) factors
in the item 1 above,
lead to an unacceptable defective behavior of the low-energy theory,
when we try to construct a model having the photon as the only
massless colourless U(1) gauge boson.
Our findings pose extremely severe constraints on such models effectively ruling them out.
One way out is to modify some of the assumptions.
We will discuss the introduction of universal extradimensions and modifications of the noncommutative field
theory at very high energy scales.

The UV/IR mixing in noncommutative theories arises from the fact that
certain classes of Feynman diagrams acquire factors of the form
$e^{i k_\mu \theta^{\mu\nu} p_\nu}$
(where $k$ is an external momentum and $p$ is a loop momentum) compared to their commutative
counter-parts.
These factors directly follow from the use of the Weyl-Moyal star-product (\ref{stardef}).
At large values of the loop momentum $p$,
the oscillations of $e^{i k_\mu \theta^{\mu\nu} p_\nu}$
improve
the convergence of the loop integrals. However, as the external momentum vanishes, $k \to 0,$
the divergence reappears and
what would have been a UV divergence is now reinterpreted as an IR divergence instead.
This phenomenon of UV/IR mixing is specific to noncommutative theories and
does not occur in the commutative settings where the physics of high energy degrees
of freedom does not affect the physics at low energies.

There are two important points concerning the UV/IR
mixing~\cite{Matusis:2000jf,Khoze:2000sy,Hollowood:2001ng,Armoni:2000xr}
which we want to stress here.
First, the UV/IR mixing occurs only in the trace-U(1) components of the
noncommutative ${\rm{U}}(N)$ theory, leaving the ${\rm{SU}}(N)$ degrees of freedom unaffected.
Second, there are two separate sources of the UV/IR mixing contributing to the
dispersion relation of the trace-U(1) gauge fields: the $\Pi_1$ effects
and the $\Pi_2$ effects, as will be explained momentarily.

A study of the Wilsonian effective action, obtained by integrating out the
high-energy degrees of freedom using the background field method,
and keeping track of the UV/IR mixing effects,
has given strong hints in favour of a non-universality in the infrared~\cite{Khoze:2000sy,Hollowood:2001ng}.
In particular, the polarisation tensor of
the gauge bosons in a noncommutative ${\rm{U}}(N)$ gauge theory takes a form
~\cite{Matusis:2000jf,Khoze:2000sy,Hollowood:2001ng}
\begin{equation}
\label{poltensor}
\Pi_{\mu\nu}^{AB} = \Pi_1^{AB}(k^2,\tilde k^2) \, \left( k^2 g_{\mu\nu} - k_\mu k_\nu \right)
+ \Pi_{2}^{AB} (k^2, \tilde k^2)\, \frac{\tilde{k}_{\mu}\tilde{k}_{\nu}}{\tilde{k}^2}
\,, \hspace{4ex} {\rm with\ } \tilde{k}_\mu = \theta_{\mu\nu} k^\nu \,.
\end{equation}
Here $A,B=0,1,\ldots N^2-1$ are adjoint labels of ${\rm{U}}(N)$ gauge fields, $A_\mu^A$,
such that $A,B=0$ correspond to the overall ${\rm{U}}(1)$ subgroup, i.e. to the
trace-U(1) factor.
The term in (\ref{poltensor})
proportional to $\tilde{k}_\mu\tilde{k}_\nu /\tilde{k}^2 $
would not appear in
ordinary commutative theories. It is transverse, but
not Lorentz invariant, as it explicitly depends on $\theta_{\mu\nu}.$
Nevertheless it is perfectly allowed in noncommutative theories.
It is known
that $\Pi_2$ vanishes
for supersymmetric noncommutative gauge theories with unbroken supersymmetry,
as was first discussed
in \cite{Matusis:2000jf}.

In general, both $\Pi_1$ and $\Pi_2$ terms in (\ref{poltensor}) are affected by the
UV/IR mixing. More precisely, as already mentioned earlier,
the UV/IR mixing affects specifically the $\Pi_1^{0\,0}$ components and
generates the $\Pi_2^{0\,0}$ components
in (\ref{poltensor}).
The UV/IR mixing in $\Pi_1^{0\,0}$ affects the running of the trace-U(1) coupling
constant in the infrared. For a pure noncommutative gauge theory In 4 continuous dimensions one finds,
\begin{equation}
\frac{1}{g(k,\tilde{k})_{{\rm{U}}(1)}^2} = \Pi_{1}^{0\,0}( k^2,\tilde k^2)
\, \rightarrow \,
-\,\frac{b_0}{(4\pi)^2} \, \log { k^2} \ , \qquad {\rm as} \
k^2\to 0
\,,
\end{equation}
leading to a logarithmic decoupling of the trace-U(1) gauge fields
from the ${\rm{SU}}(N)$ low-energy theory, see
Refs.~\cite{Khoze:2004zc,Khoze:2000sy,Hollowood:2001ng} for more detail.

For nonsupersymmetric theories,
$\Pi_2^{0\,0}$ can present more serious problems.
In theories without supersymmetry, $\Pi_2^{0\,0} \sim 1/{\tilde{k}^2},$ at small momenta,
and this leads to unacceptable quadratic IR singularities \cite{Matusis:2000jf}.
In theories with softly broken supersymmetry (i.e. with matching number of
bosonic and fermionic degrees of freedom) the quadratic singularities in $\Pi_2^{0\,0}$
cancel~\cite{Matusis:2000jf,Khoze:2000sy,Hollowood:2001ng}. However, the subleading
contribution $\Pi_2^{0\,0} \sim const,$ survives \cite{Alvarez-Gaume:2003}
unless the supersymmetry is exact.
For the rest of the paper we will
concentrate on noncommutative Standard Model candidates with softly broken supersymmetry,
in order to avoid quadratic IR divergencies. In this case,
$\Pi_2^{0\,0} \sim \Delta M^2_{\rm susy},$\footnote{$\Delta M^2_{\rm SUSY}
=\frac{1}{2}\sum_s M_s^2-\sum_f M_f^2$
is a measure of SUSY breaking.} as explained in \cite{Alvarez-Gaume:2003}.
The presence of such $\Pi_2$ effects will lead to unacceptable pathologies such as
Lorentz-noninvariant dispersion
relations giving mass to only one of the polarisations of the trace-U(1)
gauge field, leaving the other polarisation massless.

The presence of the UV/IR effects in the trace-U(1) factors
makes it pretty clear that a simple
noncommutative U(1) theory taken on its own has nothing to do with ordinary QED.
The low-energy theory emerging from
the noncommutative U(1) theory will become free at $k^2 \to 0$ (rather than just
weakly coupled) and in addition will have other pathologies
~\cite{Khoze:2004zc,Khoze:2000sy,Hollowood:2001ng,Alvarez-Gaume:2003}.
However,
one would expect that it is conceivable to embed a
commutative ${\rm{SU}}(N)$ theory, such as e.g. QCD or the weak sector of the Standard Model
into a supersymmetric noncommutative theory in the UV, but some extra care should be
taken with the QED U(1) sector~\cite{Khoze:2004zc}. We will show that
the only realistic way to embed QED into
noncommutative settings is to recover the electromagnetic U(1) from
a {\it traceless} diagonal generator of some higher ${\rm{U}}(N)$ gauge theory.
So it seems that in order to embed QED into a noncommutative theory one should
learn how to embed the whole Standard Model~\cite{Khoze:2004zc}. We will see, however,
that the additional trace-U(1) factors remaining from the noncommutative
${\rm{U}}(N)$ groups will make the resulting low-energy theories unviable
(for the 4 dimensional models considered in the first half of this paper).

In order to proceed we would like to disentangle the mass-effects due to the Higgs mechanism
from the mass-effects due to non-vanishing $\Pi_2.$ Hence we first set $\Pi_2= 0$
(this can be achieved by starting with an exactly supersymmetric theory).
It is then straightforward to show (see \cite{Jaeckel:2005wt}) that the Higgs mechanism alone cannot
remove all of the trace-U(1) factors from the massless theory.
More precisely, the following statement is true:
{\it Consider a scenario where a set of fundamental, bifundamental and adjoint Higgs fields breaks
${\rm{U}}(N_1)\times {\rm{U}}(N_2)\times\cdots \times {\rm{U}}(N_m) \rightarrow H,$
such that $H$  is non-trivial.
Then there is at least one generator of the unbroken subgroup $H$ with {\it non-vanishing trace}. This generator can be chosen such that it generates a U(1) subgroup.}

We can now count all the massless U(1) factors in a generic noncommutative theory
with $\Pi_2= 0$ and after the Higgs symmetry breaking. In general we can have the following scenarios
for massless U(1) degrees of freedom in $H$:
\begin{enumerate}[(a)\,]
\item{}\label{possa}${\rm{U}}(1)_{Y}$ is traceless and in addition there
is one or more factors of trace-U(1) in $H$.
\item{}\label{possb} ${\rm{U}}(1)_{Y}$ arises from a mixture of traceless and trace-U(1) generators of the
noncommutative product group ${\rm{U}}(N_1)\times {\rm{U}}(N_2)\times\cdots \times {\rm{U}}(N_m).$
\item{}\label{possc} ${\rm{U}}(1)_{Y}$ has an admixture of trace-U(1) generators as in (\ref{possb})
plus there are additional massless
trace-U(1) factors in $H$.
\end{enumerate}

In the following sections we will see that none of these options lead to an acceptable
low-energy theory once we have switched on $\Pi_2 \neq 0$, i.e. once we have introduced
mass differences between superpartners.
It is well-known \cite{Matusis:2000jf,Alvarez-Gaume:2003}
that
$\Pi_2 \neq 0$ leads to strong Lorentz symmetry violating effects in the dispersion relation
of the corresponding trace-U(1) vector bosons, and in particular, to mass-difference
of their helicity components. If option (\ref{possa}) was realised in nature, it would lead
(in addition to the standard photon) to a new colourless vector field with one polarisation
being massless, and one massive due to $\Pi_2.$

The options (b) and (c) are also not viable since an admixture of the
trace-U(1) generators to the photon would also perversely affect
photon polarisations and make some of them massive.

In the rest of this note we will explain these observations in more detail.

We end this section with some general comments on noncommutative Standard Modelling.
In an earlier analysis
\cite{Khoze:2004zc} the trace-U(1) factors were
assumed to be completely decoupled in the extreme infrared and, hence, were
neglected. However, it is important to keep in mind that the decoupling
of the trace-U(1)'s is
logarithmic and hence slow. For a 4 dimensional continuum theory one finds that even
in presence of a huge hierarchy between the noncommutative mass scale $M_{\rm{NC}}$,
say of the order of the Planck scale
$M_{\rm{P}}\sim 10^{19}\ \rm{GeV}$, and the scale
$\Lambda \sim (10^{-14}-10^{9})\,\rm{eV}$ (electroweak and QCD scale, respectively),
where the SU($N$) subgroup becomes strong, the ratio
\begin{equation}
\label{ration1}
\frac{g^2_{\rm U(1)}}{g^2_{{\rm{SU}}(N)}}\sim \frac{\log\left(\frac{k^2}
{\Lambda^{2}}\right)}{\log\left(\frac{M^4_{\rm{NC}}}{\Lambda^{2}k^2}\right)}
\gtrsim 10^{-3}\,
\end{equation}
is not negligible. In particular, the above inequality holds for any $M_{\rm{NC}}>k\gtrsim 2\Lambda$.
Hence, the complete decoupling of the trace-U(1) degrees of freedom at small non-zero
momenta does not appear to be fully justified
and the trace-U(1) would leave its traces in scattering
experiments at accessible momentum scales $k\sim 1\,\rm{eV}-10^{10}\,\rm{eV}$ (see Sect. 2 for more detail).

However, Eq. (\ref{ration1}) already gives us a hint how one can avoid that the trace-U(1)'s leave observable traces.
The logarithms in Eq. (\ref{ration1}) are a typical property of the 4 dimensional theory.
Adding universal extra dimensions (where gauge fields can propagate into the extra dimensions) one
expects that one gets a much faster power like
decoupling. We will explore this possibility in in Sect. \ref{powerlaw}.
Finally, starting from the original motivation from string theory another possibility to
avoid the conclusions stated above presents itself. Viewed as originating from string theory,
the noncommutative field theory is only a low energy limit.
At very high scales the noncommutative field theory is not necessarily a good description anymore.
We discuss a simple (but not too unreasonable) modification and study its consequences in Sect. \ref{birefringence}.
\section{UV/IR mixing and properties of the trace-U(1)}\label{example}
UV/IR mixing manifests itself only in the trace-U(1) part of the full noncommutative U($N$).
For this part it strongly affects
$\Pi_{1}$ and is responsible for the generation of nonvanishing $\Pi_{2}$ (if SUSY is not exact).
In this section we will briefly
review how the UV/IR mixing arises in the trace-U(1) sector and how this leads us to rule out
options (a) and (c) discussed in Sect.~1.

\subsection{Running gauge coupling}
Following Refs.~\cite{Khoze:2000sy,Hollowood:2001ng},
we will consider a U($N$) noncommutative theory with matter
fields transforming in the adjoint and fundamental representations of the gauge group.
We use the background field method,
decomposing the gauge field $A_\mu = B_\mu + N_\mu$ into a background field $B_\mu$ and a
fluctuating quantum field $N_\mu$,
and the appropriate background version of Feynman gauge,
to determine the effective action $S_{\rm eff}(B)$ by functionally integrating over the fluctuating
fields.

To determine the effective gauge coupling in the background field method,
it suffices to study the terms quadratic in the background field. In the effective action these take
the following form (capital letters denote full U($N$) indices and run from $0$ to $N^{2}-1$)
\footnote{We use euclidean momenta when appropriate and the analytic continuation when considering
the equations of motion in subsection \ref{eom}.},
\begin{equation}
S_{\rm eff}   \ni
\frac{1}{2}\int \frac{d^{4}k}{(2\pi)^4} B^{A}_{\mu}(k)B^{B}_{\nu}(-k)\Pi^{AB}_{\mu\nu}(k).
\end{equation}
At tree level, $\Pi^{AB}_{\mu\nu}=(k^2 g_{\mu\nu}-k_{\mu}k_{\nu})\,\delta^{AB}/g^2_0$ is the standard
transverse tensor originating from the gauge kinetic term. In a commutative theory, gauge and Lorentz
invariance restrict the Lorentz structure to be identical to the one of the tree level term.
In noncommutative theories, Lorentz invariance is violated by
$\theta$. The most general allowed structure is then given by Eq. (\ref{poltensor}).
The second term may lead to the strong Lorentz violation mentioned in the introduction.
This term is absent in supersymmetric theories \cite{Matusis:2000jf,Khoze:2000sy}.

Let us start with a discussion of the effects noncommutativity has on $\Pi_{1}$ and the running of the gauge coupling.
That is, for the moment, we postpone the study of $\Pi_2$-effects by considering a
model with unbroken supersymmetry\footnote{Nevertheless, we will
give general expressions for $\Pi_{1}$ valid also in the non-supersymmetric case.}.
As usual, we define the running gauge coupling as
\begin{equation}
\label{defcoupling}
\left(\frac{1}{g^{2}}\right)^{AB}=\left(\frac{1}{g^{2}_{0}}\right)^{AB}+\Pi^{AB}_{1\,\,\rm{loop}}(k).
\end{equation}
where $g^{2}_{0}$ is the microscopic coupling (i.e. the tree level contribution) and
$\Pi_{\rm{loop}}$ includes only the contributions from loop diagrams.
Henceforth, we will drop the loop subscript.

To evaluate $\Pi$ at one loop order one has to evaluate the appropriate Feynman diagrams.
The effects of noncommutativity appear
via additional phase factors $\sim \exp(i \frac{p \tilde{k}}{2})$ in the loop-integrals.
Using trigonometric relations one can group the integrals into terms where these factors combine
to unity, the so called planar parts, and those where they yield $\sim \cos ({p\tilde{k}})$,
the so called non-planar parts.

For fields in the fundamental representation, the phase factors cancel exactly\footnote{One may roughly
imagine that for each fundamental field
that appears in a Feynman diagram there is also the complex conjugate field which cancels the exponential factor.}
and only the
planar part is non-vanishing.
Fundamental fields therefore contribute as in the commutative theory \cite{Khoze:2000sy}.
In all loop integrals\footnote{To keep the equations simple we consider in this section a situation
where all particles of a given spin and
representation have equal diagonal masses.
Please note that the masses for fermions and bosons in the same representation may be different
as required for SUSY breaking.}
involving adjoint fields one finds the following factor \cite{Hollowood:2001ng},
\begin{equation}
M^{AB}(k, p) =
(-d \sin {k \tilde{p}\over 2}+ f \cos {k \tilde{p}\over 2})^{ALM}
(d \sin {k \tilde{p}\over 2}+ f \cos {k \tilde{p}\over 2})^{BML}.
\end{equation}
Using trigonometric and group theoretic relations this collapses to
\begin{equation}
M^{AB}(k,p) = - N \ \delta^{AB} (1-\delta_{0A}\cos k\tilde{p}).
\end{equation}
We can now easily see that all effects from UV/IR mixing, marked by the presence of the $\cos k\tilde{p}$,
appear only in the trace-U(1) part of the gauge group.
The planar parts, however, are equal for the U(1) and SU($N$) parts.

Summing everything up we find
the planar contribution (the coefficients $\alpha_{j},C_j,d_j$ are given in Table \ref{coefficients} and $C({\bf r})$
is the Casimir operator in the representation ${\bf r}$)
\begin{eqnarray}
\label{planarsusy2}
&&\Pi_{1\,\rm{planar}} (k^2) =
-{8 \over (4\pi )^2 }\bigg( \sum_{j, {\bf r}} \alpha_{j}  C({\bf r})
\bigg[2C_j+\frac{8}{9}d_j
\\\nonumber
&&\quad\quad\quad\quad\quad\quad\quad\quad\quad\quad\quad\quad\quad+
\int_{0}^{1} dx \left(C_j-(1-2x)^2d_j\right)\ \log {A(k^2 , x,m^2_{j, {\bf r}}) \over \Lambda^2} \bigg]\bigg),
\end{eqnarray}
where $m_{j, {\bf r}}$ is the mass of a spin $j$ particle belonging to the representation $\bf r$ of the gauge group,
\begin{equation}
A(k^2,x,m^2_{j, {\bf r}})=k^2 x(1-x)+m^2_{j, {\bf r}},
\end{equation}
and
$\Lambda$  appears via dimensional transmutation similar to $\Lambda_{\overline{\rm{MS}}}$ in QCD.
We have chosen the renormalisation scheme, i.e. the finite constants, such that $\Pi_{1\,\rm{planar}}$ vanishes
at $k=\Lambda$.
\begin{table}[!t]
\begin{center}
\begin{tabular}{|c|c|c|c|c|}
\hline  j=& scalar  & Weyl fermion & gauge boson  & ghost \\
\hline $\alpha_{j}$ & -1 & $\frac{1}{2}$ & $-\frac{1}{2}$ &1  \\
\hline  $C_j$ & 0 &  $\frac{1}{2}$& 2 & 0 \\
\hline  $d_j$ & 1 & 2 & 4 & 1 \\
\hline
\end{tabular}
\end{center}
\caption{Coefficients appearing in the evaluation of the loop diagrams.}
\label{coefficients}
\end{table}

For the trace-U(1) part the nonplanar parts do not vanish and we find
\begin{equation}
\label{pinonplanar}
\Pi_{1\,{\rm nonplanar}} = { 2\over  k^2}\left(\hat{\Pi} - \tilde{\Pi}\right) ,
\end{equation}
with
\begin{eqnarray}
\label{pi11}
\hat{\Pi} \!\!\!&=&\!\!\! {C({\bf G}) \over (4\pi)^2}\left\{
{8 d_j \over \tilde{k}^2} - k^2\left[ 12C_j - d_j\right]
\int_{0}^{1}dx \
K_{0} (\sqrt{A} |\tilde{k}|)\right\}
\ \ ,
\\
\label{pi12}
\tilde{\Pi}\!\!\!& =& \!\!\!{4C({\bf G})\over (4\pi)^2}\left\{
{ d_j \over \tilde{k}^2}-  \left(C_j k^2 -
d_j
{\partial^2 \over \partial^2 |\tilde{k}| }  \right)
 \int_{0}^{1}dx \
K_0 (\sqrt{A} |\tilde{k}|)\right\} ,
\end{eqnarray}
where $C({\bf G})=N$ is the Casimir operator in the adjoint representation.

\begin{figure}
\begin{center}
\scalebox{0.95}[0.95]{
\begin{picture}(190,180)(40,0)
\includegraphics[width=9.5cm]{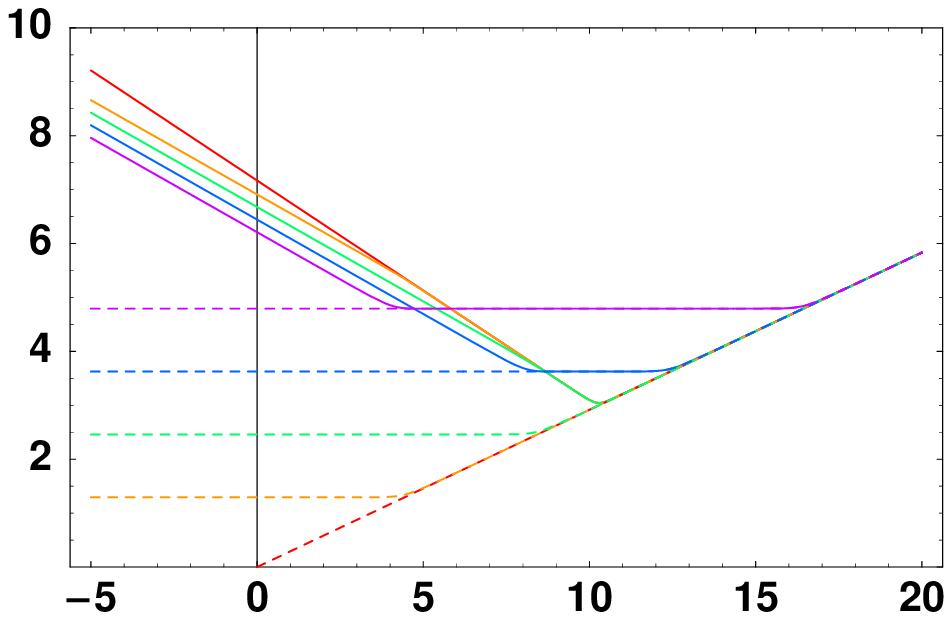}
\Text(-40,-15)[c]{\scalebox{1.2}[1.2]{$\log_{10}(k/\Lambda)$}}
\Text(-280,150)[c]{\scalebox{1.7}[1.7]{$\frac{1}{g^2}$}}
\end{picture}
}
\end{center}
\caption{The running gauge couplings $g_{\rm{U(1)}}$ (solid) and $g_{\rm{SU(2)}}$ (dashed)
for a U(2) theory with two matter
multiplets and all particles of equal mass
$m=0,10^4,10^8,10^{12},10^{16}\,\Lambda$, from top to bottom (left side, solid),
as a function of the momentum $k$, for
a choice of
$|\tilde{k}|=\theta_{\rm{eff}} |k|$, with $\theta_{\rm{eff}}=10^{-20}\Lambda^{-2}$.
}
\label{u1gaugecoupling}
\end{figure}
For illustration, we plot in Fig.~\ref{u1gaugecoupling}
the coupling (\ref{defcoupling}) for a toy model which is a supersymmetric U(2)
gauge theory with two matter multiplets and all masses (of all fields) taken to be equal.
We observe that even for large masses the running of the U(1) part (solid lines) does not stop in the infrared.
For masses smaller than the noncommutative mass scale $m^2\ll M_{\rm{NC}}$  the trace-U(1)
gauge coupling has a sharp bend at $M_{\rm{NC}}$ where the nonplanar parts
start to contribute. For larger masses the running stops at the mass scale $m^2$ only to resume running at a scale
$\sim M^{4}_{\rm{NC}}/m^2$ which is, of course,
again due to the nonplanar parts.
The dashed lines in Fig.~\ref{u1gaugecoupling} give the running of the SU(2) part which receives
no nonplanar contributions and behaves like in an ordinary commutative theory.
For $m^2=0$ the SU(2) gauge coupling reaches a Landau pole at
$k=\Lambda$, for all non vanishing masses the running stops at the mass scale.
We observe that the ratio between the SU(2) coupling and the trace-U(1) coupling is not
negligibly small over a wide range of scales,
in support of our assertion (\ref{ration1}) in Sec. 1.

Further support comes from looking at the following approximate form for the running of the gauge coupling.
We assume the hierarchy
$\Lambda^{2}\ll m^2\ll M^{2}_{\rm{NC}}$,
\begin{eqnarray}
\label{run1}
&&\!\!\!\!\!\frac{4\pi^2}{g^2_{\rm{U(1)}}}=b^{\rm{p}}_{0}\log\left(\frac{k^2}{\Lambda^2}\right),
\,\quad\quad\quad\quad\quad\quad\quad\quad\quad\quad\quad\quad\quad\quad\quad\quad\quad\,\,\rm{for}\quad
k^2 \gg M^{2}_{\rm{NC}},
\\\nonumber
&&\!\!\!\!\!\frac{4\pi^2}{g^2_{\rm{U(1)}}}=b^{\rm{p}}_{0}\log\left(\frac{k^2}{\Lambda^2}\right)
-b^{\rm{np}}_{0} \log\left(\frac{k^2}{M^{2}_{\rm{NC}}}\right),\,
\quad\quad\quad\quad\quad\quad\quad\quad\quad\,\,\, \rm{for}\quad  m^{2}\ll k^2\ll
M^{2}_{\rm{NC}},
\\\nonumber
&&\!\!\!\!\!\frac{4\pi^2}{g^2_{\rm{U(1)}}}=b^{\rm{p}}_{0}\log\left(\frac{m^2}{\Lambda^2}\right)
-b^{\rm{np}}_{0} \left[\log\left(\frac{m^2}{M^{2}_{\rm{NC}}}\right)
+\frac{1}{2}\log\left(\frac{k^2}{m^2}\right)\right],\,\quad\,\,\,\, \rm{for}\quad  k^2 \ll m^2.
\end{eqnarray}
Here, we have simplified the discussion by writing
\begin{equation}
\label{absktilde}
|\tilde{k}| =\,  M^{-2}_{\rm{NC}}\,|k|,
\end{equation}
where $ M_{\rm{NC}}$ is the noncommutativity mass-scale.
Heuristically, $M^{-2}_{\rm{NC}} \sim |\theta|$ but it may depend on the direction.
E.g., for $\theta^{\mu\nu}$ in the canonical basis,
\begin{equation}
\theta^{\mu\nu} = \left(\begin{array}{cccc} 0 & \theta_1 & 0 & 0 \\
-\theta_1 & 0 & 0 &0 \cr 0 & 0 & 0 & \theta_2 \\
0 & 0 & -\theta_2 & 0 \end{array}\right)
,
\end{equation}
only when $\theta_1 \simeq \theta_2$ one has
$M^{-2}_{\rm{NC}} = |\theta|.$ Otherwise the scale
$ M_{\rm{NC}}$ depends on $k_\mu,$
\begin{equation}
M^{-2}_{\rm{NC}} =\frac{ | \theta^{\mu\nu} k_{\nu} | }{|k|}=\,  |\theta_2|\,  \sqrt{1 +
\frac{\theta^{2}_{1}-\theta^{2}_{2}}{\theta^2_2} \frac{k_0^2+k_1^2}{k^2}}.
\end{equation}
It is nevertheless a useful scale.

The gauge coupling for the SU($N$) subgroup $g^2_{{\rm{SU}}(N)}$ is obtained by setting
$b^{\rm{np}}_{0}=0$.
For simplicity let us now consider a situation where we have only fields
in the adjoint representation.
One finds \cite{Khoze:2004zc,Hollowood:2001ng} that $b^{\rm{np}}_{0}=2b^{\rm{p}}_{0}$, and
\begin{eqnarray}
\label{run2}
&&\frac{g^{2}_{{\rm{U}}(1)}}{g^{2}_{{\rm{SU}}(N)}}=1,
\quad\quad\quad\quad\quad\quad \rm{for} \quad k^2\gg M^{2}_{\rm{NC}},
\\\nonumber
&&\frac{g^{2}_{{\rm{U}}(1)}}{g^{2}_{{\rm{SU}}(N)}}
=\frac{\log\left(\frac{k^2}{\Lambda^{2}}\right)}{\log\left(\frac{M^{4}_{\rm{NC}}}{\Lambda^{2}k^2}\right)},
\quad\,\,\,\,\, \rm{for}\quad m^{2}\ll k^2\ll M^{2}_{\rm{NC}},
\\\nonumber
&&\frac{g^{2}_{{\rm{U}}(1)}}{g^{2}_{{\rm{SU}}(N)}}
=\frac{\log\left(\frac{m^2}{\Lambda^{2}}\right)}{\log\left(\frac{M^{4}_{\rm{NC}}}{\Lambda^{2}k^2}\right)},
\quad\,\,\,\,\, \rm{for} \quad k^{2}\ll m^2.
\end{eqnarray}
To reach
\begin{equation}
\frac{g^{2}_{{\rm{U}}(1)}}{g^{2}_{{\rm{SU}}(N)}}<\epsilon = 10^{-3}
\end{equation}
we need $\log\left(\frac{M^{4}_{\rm{NC}}}{\Lambda^{2}k^2}\right)$ and in turn $M_{\rm{NC}}$ to be large.

As a generic example let
us use $\Lambda=\Lambda_{W}\sim 10^{-14}\,\rm{eV}$ (the scale where the ordinary electroweak
SU(2) would become strong, in absence of electroweak symmetry breaking)
and $k=1\,\rm{eV}$\footnote{It is obvious that $k^2\ll M^{2}_{\rm{NC}}$. In this regime our
formulas (\ref{run1}) and (\ref{run2}) approximate the full result to a very high precision since threshold effects are negligible.}. We find
\begin{equation}
\label{abschaetz}
M_{\rm{NC}}>
\Lambda^{\frac{1}{2}}k^{\frac{1}{2}}\exp\left(\frac{1}{4\epsilon}\log\left(\frac{k^2}{\Lambda^2}\right)\right)
\sim 10^{6974}\,M_{\rm{P}}.
\end{equation}
Taking electroweak symmetry breaking into account we have to replace $\log\left(\frac{k^2}{\Lambda^2}\right)$
by $\log\left(\frac{M^2_{\rm{EW}}}{\Lambda^2}\right)$ with $M_{\rm{EW}}\sim 100\,\rm{GeV}$
in (\ref{abschaetz}). We find
\begin{equation}
M_{\rm{NC}}>10^{12474}\,M_{\rm{P}}.
\end{equation}
Let us increase the coupling strength of the SU($N$) by using $\Lambda=0.5\,\rm{eV}$. $k=1\,\rm{eV}$
is now quite close to the strong coupling scale of the SU($N$). Without symmetry breaking we find
\begin{equation}
M_{\rm{NC}}>10^{131}\,M_{\rm{P}}.
\end{equation}
We might be able to reduce this number by some orders of magnitude but without using an extreme field content
it remains always extraordinarily large. Indeed, one can typically find a scale $k$ which is not too close to the strong coupling scale of the SU($N$) which
strengthens the bounds dramatically. Therefore, as a conservative estimate we
propose\footnote{Please note that this implies small vacuum expectation value for the B-fields that could be the origin of noncommutativity in string theory. The reason is that
$M^{-2}_{\rm{NC}}\sim \theta\sim \frac{1}{const+B}B\frac{1}{const-B}$ and hence $M_{\rm{NC}}\to\infty$ for $B\to 0$ (we have omitted the Lorentz indices for simplicity).}
\begin{equation}
M_{\rm{NC}}>10^{100}\,M_{\rm{P}}.
\end{equation}

Let us note that this strong constraint is based on the assumption that
the 4 dimensional noncommutative field theory is a valid description up
to arbitrarily high momentum scales. This assumption is not necessarily fulfilled if the noncommutative theory is
embedded into a more fundamental theory, e.g. string theory. In the later Sects. \ref{powerlaw} and \ref{birefringence}
we will investigate situations where this assumption is not valid anymore and the constraints can be weakened.

\begin{figure}[t]
\begin{center}
\scalebox{0.9}[0.9]{
\begin{picture}(190,110)(0,0)
\SetOffset(3,10)
\ArrowLine(40,60)(0,100)
\ArrowLine(0,20)(40,60)
\DashLine(40,60)(100,60){2}
\ArrowLine(140,100)(100,60)
\ArrowLine(100,60)(140,20)
\Vertex(40,60){2}
\Text(20,60)[c]{\scalebox{1.0}[1.0]{$g(k)$}}
\Text(120,60)[c]{\scalebox{1.0}[1.0]{$g(k)$}}
\Vertex(100,60){2}
\Text(70,75)[c]{\scalebox{1.1}[1.1]{$k$}}
\Text(70,65)[c]{\scalebox{1.2}[1.2]{$\longrightarrow$}}
\end{picture}}
\end{center}
\vspace{-1.0cm}
\caption{A typical Feynman diagram for scattering. The effective coupling $g$ depends on the momentum $k$.}
\label{scattering}
\end{figure}
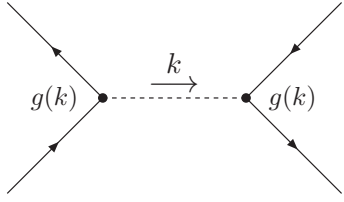

To conclude this subsection, let us point out that, in a scattering experiment (as depicted in Fig. \ref{scattering}),
$k$ is really the scale of the internal momentum, and therefore, non-vanishing.
$\tilde{k}$, too, is non-vanishing in appropriate (remember that we have Lorentz symmery violation)
directions of $t$-channel scattering.

\subsection{The effects of a non vanishing $\Pi_{2}$ from SUSY breaking}\label{eom}

In the previous subsection we made
$\Pi_{2}$ vanish by working in a supersymmetric theory.
Let us now study, what happens, when supersymmetry is (softly) broken.

Looking only at the trace-U(1) degrees of freedom of a generic noncommutative theory we have
\begin{equation}
\Pi_{2}=2\sum_{j}\alpha_{j}\left[(3\tilde{\Pi}_{j}-\hat{\Pi}_{j})\right].
\end{equation}
One easily checks that
\begin{equation}
\Pi_{2}\sim \sum_{j} \alpha_{j} d_j f(k^2,\tilde{k}^2,m_j).
\end{equation}
If SUSY is unbroken, all masses are equal. Using supersymmetric matching between
bosonic and fermionic degrees of freedom,
\begin{equation} \label{susycanc}
\sum_{j} \alpha_{j} d_j=0,
\end{equation}
we reproduce the vanishing of $\Pi_{2}$.
If SUSY is softly broken this cancellation is not complete anymore
(in fact (\ref{susycanc}) still holds and this removes the leading power-like IR divergence
in $\Pi_{2}$, however, the subleading effects in $\Pi_{2}$ survive).
$\Pi_{2}$
gets a contribution \cite{Alvarez-Gaume:2003}
\begin{eqnarray}
\label{pi2}
\Pi_{2}\!\! &=& \!\!D\sum_{j}\alpha_{j}d_jm^{2}_{j}\left[K_{0}(m\tilde{k})+K_{2}(m\tilde{k})\right]+O(k^2)
\\\nonumber
\!\! &=& \!\! C \Delta M^2_{\rm{SUSY}}+C'\sum_{j}\alpha_{j}d_j m^{2}_{j}\log(m^2_{j}\tilde{k}^2)+\cdots,
\end{eqnarray}
with known constants $C$, $C'$ and $D$.
This has dire consequences for the gauge boson. Let us look at the equations of motion resulting from this
additional Lorentz symmetry violating contribution to the polarisation tensor.

In presence of a Higgs field which generates a mass term $m^2$ and using unitary gauge the field
equations in presence of non vanishing $\Pi_2$ read
\begin{equation}
\label{fieldnc} \left(\Pi_{1}(k^{2}g_{\mu\nu}-k_{\mu}k_{\nu})
+\Pi_{2}\frac{\tilde{k}_{\mu}\tilde{k}_{\nu}}{\tilde{k}^{2}}-m^{2}g_{\mu\nu}\right)A^{\nu}=0.
\end{equation}
Using that unitary gauge implies Lorentz gauge,
$k_{\mu}A^{\mu}=0$, we can simplify
\begin{equation}
\label{fieldnc2}
(\Pi_{1}k^{2}-m^{2})A_{\mu}+\Pi_{2}\frac{\tilde{k}_{\mu}\tilde{k}_{\nu}}{\tilde{k}^{2}}A^{\nu}=0.
\end{equation}
To proceed further it is useful to specify a direction for the
momentum and the noncommutativity parameters. The photon flies in
3-direction and we have
\begin{equation}
k^{\mu}=(k^{0},0,0,k^{3}).
\end{equation}
What is the corresponding value of $\tilde k$?
Since $\theta^{\mu\nu}$ breaks Lorentz invariance, we
need to specify $\theta^{\mu\nu}$ in a particular frame.
For the latter, a natural one is the system where the cosmic microwave background is at rest.
In this frame, we assume that the only non-vanishing components of $\theta^{\mu\nu}$ are
\begin{equation}
\label{thetamatrix}
\theta^{13}=-\theta^{31}=\theta .
\end{equation}

This yields,
\begin{equation}
\tilde{k}_{\mu}=\theta_{\mu\nu}k^{\nu}=(0,\theta k^{3},0,0),\quad
|\tilde{k}^{\, 2}|=(\theta k^{3})^{2}.
\end{equation}
We start with the ordinary transverse components of $A^{\nu}$,
\begin{equation}
A^{\nu}_{1}=(0,1,0,0).
\end{equation}
In this direction, (\ref{fieldnc2}) yields
\begin{equation}
\label{direction1}
(\Pi_{1}k^{2}-m^{2}-\Pi_{2})A_{1,\nu}=0.
\end{equation}
In the other transverse direction,
\begin{equation}
A^{\mu}_{2}=(0,0,1,0),
\end{equation}
we find
\begin{equation}
(\Pi_{1}k^{2}-m^{2})A_{2,\nu}.
\end{equation}
Finally we have the third polarisation (which can be gauged away if and only if $m^2=0$),
\begin{equation}
A^{\mu}=(a,0,0,b),\quad k^{0}a-k^{3}b=0
\end{equation}
which results in
\begin{equation}
(\Pi_{1}k^{2}-m^{2})A_{3,\nu}.
\end{equation}
We note that the different polarisation states do not
mix due to the presence of $\Pi_{2}$. The second and the third polarisation state behave more
or less like in the ordinary commutative case. However, the first has a modified
equation of motion, (\ref{direction1}),
in presence of a non-vanishing $\Pi_{2}$\footnote{One might argue that instead of
Eq. (\ref{direction1}) one has to use the rescaled equation (we set $m^2=0$ for
simplicity)
$k^2-\frac{\Pi_{2}(k^2,\tilde{k}^2)}{\Pi_{1}(k^2,\tilde{k^2})}=0$.
For $k^2\to 0$, the second term vanishes since $\Pi_{1}$ diverges in this limit.
Therefore, we find an additional solution.
However, this solution is rather strange. It does not correspond to a pole in the
propagator (it goes like a $\log$).
Moreover, if one calculates the cross section $\Pi_{2}$ still upsets the angular
dependence quite severely compared to
the ordinary commutative case.}.

This is another strong argument against a trace-U(1) being the photon \cite{Alvarez-Gaume:2003}.
If the gauge symmetry is unbroken and $m^2=0$ we usually have two massless polarisations.
However, a non vanishing $\Pi_{2}$ reduces this to one. The other one gets an
additional mass $\case{\Pi_{2}}{\Pi_{1}}$.
Since only one polarisation is affected this is a strong Lorentz symmetry violating effect.
Moreover, a negative $\Pi_{2}$ would lead to tachyons while a positive mass is phenomenologically ruled
out by the constraint \cite{Eidelman:2004wy}
\begin{equation}
m_{\gamma}<6\times 10^{-17}\,\rm{eV}
\end{equation}
on the photon mass\footnote{Even fine-tuning of (\ref{pi2}) to zero is not an option.
Since we have only a finite number of masses this is at best possible for a finite number of values
of $|\tilde{k}|$ and we will surely find values of $|\tilde{k}|$ where $\Pi_{2}$ is nonzero.}.

If we take the trace-U(1) as an additional (to the photon)
gauge boson from the unbroken subgroup $H$, we would still get strong Lorentz symmetry
violation since the trace-U(1) is not completely decoupled.

In summary, we found in this section that additional trace-U(1) subgroups
are not completely decoupled and should lead to observable effects.
In particular, if SUSY is not exact we have non-vanishing $\Pi_{2}$
which gives rise to strong Lorentz symmetry violation which has not been observed.
This rules out possibilities (\ref{possa}) and (\ref{possc}) of Sec.~1.
Moreover, we confirmed that a trace-U(1) is not suitable as a photon candidate.

\section{Mixing of trace and traceless parts}\label{u2example}
{}From the previous section we concluded that the trace-U(1) groups are unviable
as candidates for the SM photon.
Therefore, it has been suggested to construct the photon from traceless U(1) subgroups \cite{Khoze:2004zc}.
It turns out, however, that typically trace and traceless parts mix and the trace parts
contribute their Lorentz symmetry violating properties
to the mixed particle.

For U(2) broken by a fundamental Higgs, the standard Higgs mechanism yields the symmetry breaking $U(2)\to U(1)$.
However, the remaining U(1) is a mixture of trace and traceless parts.
If SUSY is broken, the trace-U(1) has a $\Pi_{2}$ part in the polarisation tensor.
Taking this into account we find the following matrix for the equations of motion
\begin{eqnarray}
\label{final}
\begin{tiny}
\left(\begin{array}{cccccc}
 \Pi^{\rm{U(1)}}_{1}k^{2}-\Pi_{2}-m^{2} &  m^2 &  &  &  &  \\
m^2 & \Pi^{\rm{SU(2)}}_{1}k^2-m^{2} &  &  &  &  \\
 &  & \Pi^{\rm{U(1)}}_{1}k^{2}-m^{2} & m^2 &  &  \\
 &  &  m^2 &\Pi^{\rm{SU(2)}}_{1}k^2-m^{2}   &  &  \\
 &  &  &  & \Pi^{\rm{U(1)}}_{1}k^2-m^{2} &m^{2}  \\
 &  &  &  & m^{2} &\Pi^{\rm{SU(2)}}_{1}k^2-m^{2}
\end{array}  \right),
\end{tiny}
\end{eqnarray}
where the adjoint U(2) and polarisation indices are $(0,1),(3,1),(0,2),(3,2),(0,3),(3,3)$.
We omitted the values $1$ and $2$ for the adjoint U(2) indices which do not mix with
the trace-U(1) and are not qualitatively different from the commutative case.

The matrix is block
diagonal and the second and third polarisation (lower right corner)
behave more or less like their commutative counterparts.
We can concentrate on the upper left $2\times2$ matrix corresponding
to the transverse polarisations affected
by $\Pi_{2}$.

This $2\times2$ matrix admits two solutions for the equations of motion. Expanding for small $\Pi_{2}$ we
find,
\begin{eqnarray}
\label{u2solution}
\left(\Pi^{\rm{U(1)}}_{1}+\Pi^{{\rm{SU}}(N)}_{1}\right)k^2\!\!&=&\!\!\Pi_{2}+O(\Pi^{2}_{2}),
\\\nonumber
\left(\Pi^{\rm{U(1)}}_{1}+\Pi^{{\rm{SU}}(N)}_{1}\right)k^2\!\!&=&\!\!\
\frac{\left(\Pi^{\rm{U(1)}}_{1}+\Pi^{{\rm{SU}}(N)}_{1}\right)^2}{\Pi^{{\rm{U}}(1)}_{1}\Pi^{{\rm{SU}}(N)}_{1}} m^2 +\frac{\Pi^{{\rm{SU}}(N)}_{1}}{\Pi^{{\rm{U}}(1)}_{1}}\Pi_{2}+O(\Pi^{2}_{2}),
\end{eqnarray}
in analogy to (\ref{direction1}). In absence of $\Pi_{2}$ the first solution in
Eq. (\ref{u2solution}) is a massless one
corresponding to the massless combination of gauge bosons (think of it as the photon).
The second is a massive combination (similar to the $Z$ boson). The presence of non-vanishing $\Pi_{2}$
again leads to a mass $\frac{\Pi_{2}}{\Pi^{\rm{U(1)}}}$ for the first
solution and rules out the ``massless'' combination
as a reasonable photon candidate.

This example demonstrates that the disastrous effects of $\Pi_{2}$ are also present in any combination
which has an admixture of trace-U(1) degrees of freedom.
Hence, this rules out possibilities (\ref{possb}) and (\ref{possc}) from the introduction.

\section{Universal extra dimensions and power law running in the UV and IR}\label{powerlaw}
In the introduction we already mentioned that a possible way out of the dilemma with the trace-U(1)'s is
the introduction of universal extra dimensions\footnote{A particularly interesting possibility is that the extra dimensions may arise dynamically \cite{Aschieri:2006uw}.}. Let us now investigate this option.

In most of the following discussion we will adopt a four-dimensional point of view
in describing extra-dimensional theories. That is, because we are interested
in renormalisation group effects associated with the 4-dimensional momentum, it makes more sense to
include the effects of extra dimensions by considering the effect of a simple Kaluza-Klein tower of states.
(In the UV-complete string models there are other effects which, at one-loop order and in compact dimensions
significantly larger than the string length, will be secondary.)

Intuitively it is obvious that the main factor affecting the running of the gauge couplings will be the
noncommutativity parameter $\tilde{k}$, and in particular how it mixes the additional (compact)
dimensions
with the ordinary four large dimensions.
We will now give a somewhat heuristic presentation of how $\tilde{k}$ affects the running of the gauge couplings.
A more
precise and general calculation is given in \cite{AJKR} and we will just quote
the results from there in the last part of this section.

\subsection{The UV regime}

Let us start by briefly reviewing power law running in the UV at scales well
above the compactification scale. In the UV regime the planar diagrams
dominate the two point function and so
there is no difference to the ordinary commutative case (see \cite{Dienes:1998vg}).
Because of this
it is sufficient to use an intuitive approach based on thresholds\footnote{A fuller treatment based on
dimensional regularisation is presented in \cite{AJKR}. An even better one
is presented in Ref.~\cite{ghilencea}. In those treatments it becomes
evident that higher-dimensional operators appear in the effective action.
These operators are due to a different form of UV/IR mixing from
regions of KK momenta that are zero in some directions and high in others.
These difficulties are absent for the IR regime which is the main point
of interest in the present discussion so we do not dwell on them here.}.

Consider first the most simple case of one compact extra dimension
of size $M^{-1}_{c}$. Neglecting threshold effects
the one loop running of the gauge coupling in four dimensions typically follows
($t=\log(k)$)
\begin{equation}
\frac{\partial}{\partial t}g^2=\sum_{m^{2}_{i}<k^2}c_{i} g^4,
\end{equation}
where the $c_{i}$ are coefficients depending on the spin and representation of the particle $i$. In the sum only
particles with mass $m^{2}_{i}$ smaller than the momentum scale $k^2$ contribute (in any suitable
massive renormalisation scheme).
This leads to the typical decoupling
of massive modes.
For simplicity, let us now consider a situation where all particles have (approximately) the same mass $m^2\ll M^{2}_{c}$.
We find
\begin{equation}
\frac{\partial}{\partial t}g^2=-b_0 g^4,\quad \rm{for}\quad m^2\ll k^2\ll M^{2}_{c},
\end{equation}
where we have chosen the sign of the constant $b_0$ such that it is positive when the theory is asymptotically free.
(For example, in ${\cal N}=2$ supersymmetric pure gauge theory $b_0 =  N /(4\pi^2)$ in this notation.)

Above the compactification scale, more precisely at $m^2+M^{2}_{c}<k^2< m^2+4 M^{2}_{c}$,
the first Kaluza-Klein mode gives an identical contribution to the $\beta$-function, and in general one finds
\begin{equation}
\label{flow0}
\frac{\partial}{\partial t}g^2=-N_{\rm{KK}}(k)b_0 g^4,
\end{equation}
where $N_{\rm{KK}}(k)$ is the number of Kaluza-Klein modes (including the zero mode)
contributing at the scale $k$. Since the mass of the $n$th Kaluza-Klein mode is given by
$\sqrt{m^2+n^2M^{2}_{c}}$ one easily finds the approximate formula
\begin{equation}
\label{number}
N_{\rm{KK}}(k)\approx C_{1} \frac{k}{M_{c}} \quad\rm{for}\quad k\gg M_{c},
\end{equation}
where we have introduced the constant $C_{1}$ to account for the details of the compactification
and threshold effects.
This already suggests power law running. More precisely,
one easily checks that for $k^2\gg M^{2}_{c}$ and appropriate initial conditions the solution approaches
\begin{equation}
\label{powersolution}
g^{2} \approx \frac{1}{C_{1} b_0}\frac{M_{c}}{k}
\end{equation}
which is indeed a power law.

Expressions (\ref{number}) and (\ref{powersolution}) are easily generalized to arbitrary dimension
$D=n+4$ ($k^2\gg M^{2}_{c}$)
\begin{eqnarray}
\label{flow}
\frac{\partial}{\partial t}g^2\!\!&=&\!\!-N_{\rm{KK}}(k)b_0 g^4,\\\nonumber
 N_{\rm{KK}}(k)\!\!&\approx&\!\!  C_{n}\left(\frac{k}{M_{c}}\right)^{n} ,\\\nonumber
 g^{2}\!\!&\approx&\!\!\frac{n}{C_{n}b_0}\left(\frac{M_{c}}{k}\right)^{n},
\end{eqnarray}
where again the constant $C_{n}$ depends on the details of
the compactification.

The flow equation (\ref{flow}) for the running coupling can be also discussed
using the more natural effective coupling $\hat{g}^2$ of the $D$-dimensional theory,
\begin{equation}
\label{rescaling}
\hat{g}^2=\left(\frac{k}{M_{c}}\right)^{n}g^2.
\end{equation}
From the lower-dimensional viewpoint
(\ref{rescaling}) can be understood by remembering that the amplitudes of all Kaluza-Klein modes add up and therefore
increase the effective coupling by a factor $N_{\rm{KK}}$.
Inserting (\ref{rescaling}) into (\ref{flow}) yields the flow equation for $\hat{g}^{2}$,
\begin{equation}
\label{rescaled}
\frac{\partial}{\partial t}\hat{g}^2=n\hat{g}^{2}-C_{n}b_{0}\hat{g}^{4}
=(n-C_{n}b_0\hat{g}^{2})\hat{g}^{2}, \quad\rm{for}\quad k^2\gg M^{2}_{c}.
\end{equation}
If we start at small values for $\hat{g}^2$ the coupling increases toward the infrared until it reaches a fixed
point at $\hat{g}^2_{\rm fixed}=\frac{n}{C_{n}b_0}$.
The corresponding coupling of the 4-dimensional theory is then
\begin{equation}
\label{fixed4d}
g^2_{\rm fixed} (k) =\hat{g}^2_{\rm fixed} \left(\frac{M_{c}}{k}\right)^{n}=
\frac{n}{C_{n}b_0}\left(\frac{M_{c}}{k}\right)^{n},
\end{equation}
in agreement with the last equation in (\ref{flow}).
This discussion implies that power-law running in extra dimensions originates
from a fixed point in the effective higher-dimensional coupling constant $\hat{g}^2$.
This implies that the power-law running of $g^2$ is a strong coupling phenomenon in terms
of $\hat{g}^2$
and one should exercise caution since Eqs. (\ref{flow}) and (\ref{rescaled}) are one-loop results.
In particular a large number of extra-dimensions increases the value of the fixed point coupling
and the approximation may break down. The issues of existence of a fixed point
of $\hat{g}^2$ were investigated in literature on extra-dimensional
gauge theories, see e.g. \cite{Gies:2003ic}.

From now on we will continue assuming that (ordinary commutative) extra-dimensional
gauge theories do provide a power-law running of the coupling in the extreme ultraviolet
(i.e. at energies well above the compactification scale).
We will then show that in noncommutative settings the mixing between ultraviolet and infrared
degrees of freedom will induce in the extreme infrared a power-law decoupling of the trace-U(1) degrees
of freedom.

\subsection{IR running -- noncommutativity restricted to 4 dimensions}\label{simple}

As specified in Eq. (\ref{poltensor}) $\Pi_{1}$ and therefore the gauge coupling
depends on the additional scale $\tilde{k}$
(cf. \cite{Matusis:2000jf,Minwalla:1999px,Hollowood:2001ng,Khoze:2000sy})
$\tilde{k}^{\mu}=\theta^{\mu\nu}k_{\nu}$.
In fact, the coupling depends only on the absolute values $|\tilde{k}|$ as well as
$|k|$, as can be seen from Eqs. (\ref{pi11}) and (\ref{pi12}).

Since we are mostly interested in low-energy physics (compared to the compactification scale)
the effects of extra dimensions can contribute only through loops in perturbation theory.
Thus the external momenta $k_\mu$ are taken to be 4-dimensional,
i.e. external particles will not include excited Kaluza-Klein modes, while
internal loop momenta $p_\mu$ (in Feynman diagrams) are kept general.

In this section we consider a scenario where
only the four infinite dimensions are noncommutative,
\begin{equation}
\label{fourdcase}
\theta^{\mu\nu} \neq 0, \quad \theta^{\mu b}=0, \quad
\theta^{a b}=0
\end{equation}
where
$\mu,\nu=0,\ldots,3$ and $a, b=4,\ldots,3+n.$

From Eq. (\ref{defcoupling}) together with (\ref{pinonplanar}) one easily finds
that in a 4-dimensional noncommutative gauge theory with all particles of equal
non-zero mass $m$, the trace-U(1)
couplings runs according to
\begin{equation}
\label{irrun}
\frac{\partial}{\partial t}g^2=b^{\rm{np}}_{0} g^4
\quad\rm{for} \quad k^2\ll \min\left(M^{2}_{\rm{NC}},\frac{M^{4}_{\rm{NC}}}{m^2}\right).
\end{equation}
Here $b^{\rm{np}}_{0}$ is a positive number
which specifies the non-planar contribution to the running gauge coupling.

From Eq. (\ref{irrun}) one can see that in general noncommutative theory when we lower momentum-scale $k^2$ sufficiently,
even very massive modes start to contribute. This holds for Kaluza-Klein modes, too, as long as
we have noncommutativity only in the four infinite
dimensions according to Eq. (\ref{fourdcase}).
In analogy to (\ref{flow}) we find
($k^2\ll\min(M^{2}_{\rm{NC}},\frac{M^{4}_{\rm{NC}}}{M^{2}_{c}})$)
\begin{eqnarray}
\label{flowir}
\frac{\partial}{\partial t}g^2\!\!&=&\!\! N^{\rm{IR}}_{\rm{KK}}(k)b^{\rm{np}}_0 g^4,
\\\nonumber
 N^{\rm{IR}}_{\rm{KK}}(k)\!\!&\approx&\!\!  C^{\rm{IR}}_{n}\left(\frac{M^{2}_{\rm{NC}}}{M_{c}k}\right)^{n} ,
 \\ \nonumber
 g^{2}\!\!&\approx&\!\!\frac{n}{C^{\rm{IR}}_{n}b^{\rm{np}}_0}\left(\frac{kM_{c}}{M^{2}_{\rm{NC}}}\right)^{n}.
\end{eqnarray}
The right hand side of the IR flow equation in (\ref{flowir}) has the opposite sign
to that of the UV flow equation (\ref{flow}). This implies that the
trace-U(1) coupling $g^2$ becomes small in the IR and the UV regimes. The enhancement by
the $N^{\rm{IR}}_{\rm{KK}}(k)$ factor gives the {\em power-like} decoupling
of these unwanted degrees of freedom from the SU($N$) theory (which is unaffected by the UV/IR mixing effects).

\subsection{IR running for arbitrary noncommutativity}

If the matrix
$\theta^{\mu\nu}$ has nonvanishing entries that mix the ordinary four dimensions with the
extra dimensions we may have a non-vanishing
\begin{equation}
\hat{k}^{a}  =  \theta^{a\nu}k_{\nu}\,\,\,\,\,\,\,(a=4\ldots,3+n).
\end{equation}
In the calculation of the polarisation tensor this
leads to phase factors in the sum over the Kaluza-Klein modes,
\begin{equation}
\sum_{m\in\mathbb{Z}^{n}} e^{i\frac{m}{R}\cdot\hat{k}}
\end{equation}
(in addition to the usual $\theta$-dependent phases
in non-planar contributions).
In this situation it is advantageous to directly perform the sum
over Kaluza-Klein modes in the polarisation tensor.
We have done
this explicitly in \cite{AJKR}.
Here we will quote the result (for an ${\mathcal{N}}=2$ supersymmetric U($N$) theory
without adjoint matter fields),
\begin{equation}
\label{resultshort}
\Pi_{1}
=const+2\frac{C(\mathbf{G})}{(4\pi)^{2}}(4\pi)^{\frac{n}{2}}\Gamma\left(\frac{n}{2}\right) \prod_{i}R_{i}
\left(|\tilde{k}|^{-n}\right),
\end{equation}
where $R_{i}$ are the compactification radii and $\tilde{k}$ is now the total
noncommutative momentum $\tilde{k}^{M}=\theta^{M\nu}k_\nu $ ($M=0\ldots 3+n$).
This equation is valid for 
\begin{equation}
\label{regimelow}
k\ll \min\left(M_{c},\frac{M^{2}_{\rm{NC}}}{M_{c}}\right),
\end{equation}
with $M_{\rm{NC}}$ still defined as $M_{\rm{NC}}^{-2} = \frac{ |\tilde k | }{|k|}$.

The fact that the actual running
is now given by replacing the 4-dimensional components of
$\tilde{k}$ with the total ${\tilde k}$ is not too surprising since
the infrared running comes from very ultraviolet modes, i.e. it involves momenta much higher than the
compactification scale where the theory is effectively higher-dimensional. At these scales there is no
distinction between the ordinary four dimensions and the extra dimensions.

Eq. (\ref{resultshort}) has the additional
advantage that it already corresponds to the integrated result. It directly gives $g(k)$ without
the need to solve a differential equation
($R_{i}=1/M_{c}$),
\begin{equation}
\label{resultu1}
g^{2}_{{\rm{U}}(1)}(k)
=\frac{1}{A_{{\rm{U}}(1)}+\frac{C^{\rm{IR}}_{n}b^{\rm{np}}_{0}}{n}\left(\frac{M^{2}_{\rm{NC}}}{M_{c}k}\right)^{n}}.
\end{equation}
Here we have fixed,
\begin{eqnarray}
C^{\rm{IR}}_{n}\!\!&=&\!\! \frac{n}{2}(4\pi)^{\frac{n}{2}}\Gamma\left(\frac{n}{2}\right),
\\\nonumber
b^{\rm{np}}_{0}\!\!&=&\!\! \frac{4}{(4\pi)^{2}}C(\mathbf{G})
,
\end{eqnarray}
where we still consider the ${\mathcal{N}}=2$ case and none of the matter fields are in the
adjoint representation\footnote{A generalisation to an arbitrary number of matter multiplets
can be easily obtained from the results given in \cite{AJKR}.}.
$A_{{\rm{U}}(1)}$ is a renormalisation constant determined from the bare coupling and the planar diagrams only.
Therefore in the regime (\ref{regimelow}) this constant is connected to the gauge coupling
of the SU($N$)-part (up to logarithmic corrections which we neglected in our approximation)
\begin{equation}
\label{resultsun}
g^{2}_{{\rm{SU}}(N)}(k)\approx\frac{1}{A_{{\rm{SU}}(N)}}\quad\rm{with}\quad A_{{\rm{U}}(1)}=A_{{\rm{SU}}(N)}.
\end{equation}

\subsection{Lorentz violating mass term for trace-U(1)}\label{massec}
In noncommutative field theories the gauge coupling is not the only part of the polarisation tensor
that is affected by power law running. Recall that in noncommutative field
theories the (4-dimensional) polarisation tensor has
an additional Lorentz symmetry violating part \cite{Matusis:2000jf,Khoze:2000sy}, which is
called $\Pi_{2}$ in Eq. (\ref{poltensor}).

For softly broken supersymmetry only the IR-singular (pole) contribution to $\Pi_2$ vanishes,
but a constant term
\begin{equation}
\label{nonsusy}
\Pi_{2}\sim\Delta M^{2}_{\rm{SUSY}},\quad
\Delta M^2_{\rm SUSY}=\frac{1}{2}\sum_s M_s^2-\sum_f M_f^2,
\end{equation}
remains.
In (\ref{nonsusy}) the sums run over \emph{all} bosons and fermions. Therefore, if we have compactified
extra dimensions, we must include the Kaluza-Klein modes, effectively multiplying the four-dimensional
$\Delta M^{2}_{\rm{SUSY}}$ by the number of Kaluza-Klein modes.
The number of contributing Kaluza-Klein modes is, again, given roughly by $N^{\rm{IR}}_{\rm{KK}}$
of Eq. (\ref{flowir}).
Hence, we find
\begin{equation}
\label{pi2}
\Pi_{2}\sim N^{\rm{IR}}_{\rm{KK}}(k)\Delta M^{2}_{\rm{SUSY}}\sim
\left(\frac{M^{2}_{\rm{NC}}}{M_{c}k}\right)^{n}\quad\rm{for}
\quad k^{2}\ll \min(M^{2}_{\rm{NC}},\frac{M^{4}_{\rm{NC}}}{M^{2}_{c}}).
\end{equation}

Repeating the analysis of Sect. \ref{eom} one finds, again, one ordinary massless polarisation state
and one with  a Lorentz symmetry breaking mass,
\begin{equation}
m^{2}_{\rm{LV}}\sim \frac{\Pi_{2}}{\Pi_{1}}\sim \Delta M^{2}_{\rm{SUSY}},
\end{equation}
which is roughly constant although both $\Pi_{1}$ and $\Pi_{2}$ scale with a power law. Yet, these power laws
cancel since they are the same for $\Pi_{1}$ and $\Pi_{2}$.

\subsection{Weaker constraints from power law running}\label{bounds}
We found in Sect. \ref{massec} that the Lorentz violating mass term for the trace-U(1) factors
remains roughly constant. Hence trace-U(1)'s are still unsuitable as photon candidates. With a similar argument
as in Sect. \ref{u2example} one finds that this also holds for mixtures of trace and traceless parts.
Therefore a suitable photon candidate must be constructed (as in four dimensions)
from an unbroken combination of traceless generators. In \cite{Jaeckel:2005wt}
we found that such a combination can only exist together with
additional unbroken U(1)'s which have nonvanishing trace.
Here the results of Sect.~\ref{powerlaw} help us, since they allow for a fast decoupling
of trace-U(1) degrees of freedom. This is in contrast to the four-dimensional case, where the (only)
logarithmic decoupling necessitated incredibly large noncommutativity scales
$M_{\rm{NC}}\gg M_{\rm{P}}$.
With additional (compactified) space dimensions we have power law running according to (\ref{flowir}).
This decouples the unwanted trace-U(1)'s much faster in the IR thereby weakening the constraints dramatically.

Let us now estimate the new constraints obtained from power law running.
As already mentioned earlier, current experiments probe the regime well below $M_{c}$.
To apply Eq. (\ref{resultu1}) we also need $k\ll k_{s}$,
\begin{equation}
k_{s}=\frac{M^{2}_{\rm{NC}}}{M_{c}}.
\end{equation}
This is also assured, since the discussion of Sect. \ref{simple}
shows that for $k\sim k_{s}$ the trace-U(1) and the SU($N$) have gauge couplings which are of the same order.
(Until $k\sim M_{\rm{NC}}$ both gauge couplings are approximately equal and
power law running sets in only below $k_{s}$.)

\begin{figure}
\begin{center}
\scalebox{0.90}[0.90]{
\begin{picture}(190,180)(40,0)
\includegraphics[width=9.5cm]{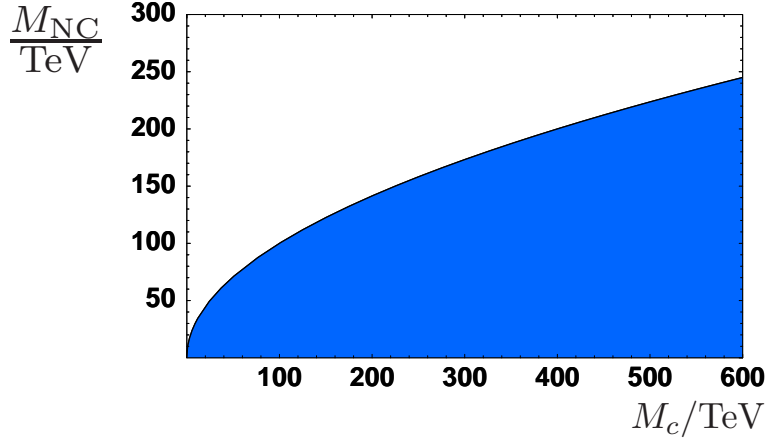}
\Text(-30,-10)[c]{\scalebox{1.4}[1.4]{$M_{c}/\rm{TeV}$}}
\Text(-300,150)[c]{\scalebox{2.0}[2.0]{$\frac{M_{\rm{NC}}}{\rm{TeV}}$}}
\end{picture}
}
\end{center}
\caption{Excluded regions in the $(M_{c},M_{\rm{NC}})$-plane
(in TeV). The blue region is excluded because the
trace-U(1) still has nonnegligible coupling.
We have chosen
$\epsilon=0.05$, $C_{1}b^{\rm{np}}_{0}=0.1$,
$g^{2}_{{\rm{SU}}(N)}(k_{0})=0.2$, $k_{0}=0.1\,\rm{TeV}$, and
$n=1$.} \label{exclusion}
\end{figure}

Neglecting the slow logarithmic running of the SU($N$) couplings we find from Eqs. (\ref{resultu1}) and (\ref{resultsun}),
\begin{eqnarray}
\frac{g^{2}_{{\rm{U}}(1)}}{g^{2}_{{\rm{SU}}(N)}}
\!\!&\approx&\!\! \frac{n}{C^{\rm{IR}}_{n}b^{\rm{np}}_{0}}\frac{1}{g^{2}_{{\rm{SU}}(N)}(k_{s})}
\left(\frac{k}{k_{s}}\right)^{n}
= D k^{n}\left(\frac{M_{c}}{M^{2}_{\rm{NC}}}\right)^{n} \quad\rm{for} \quad k\ll k_{s}\\\nonumber
D\!\!&=&\!\!\frac{n}{C^{\rm{IR}}_{n}b^{\rm{np}}_{0}}
\frac{1}{g^{2}_{{\rm{SU}}(N)}(k_{s})}\sim \frac{(4\pi)^{2}}{4 N g^{2}_{{\rm{SU}}(N)}},
\end{eqnarray}
where the $\sim$ in the second line holds for a pure noncommutative U($N$).
To have
\begin{equation}
\label{bound}
\frac{g^{2}_{{\rm{U}}(1)}(k_{0})}{g^{2}_{{\rm{SU}}(N)}(k_{0})}<\epsilon
\end{equation}
we need
\begin{equation}
\label{const1}
\frac{M^{2}_{\rm{NC}}}{M_{c}}>k_{0}\left(\frac{D}{\epsilon}\right)^{\frac{1}{n}}.
\end{equation}
As an illustration we have plotted the excluded region in Fig. \ref{exclusion}.
This shows that when we allow for a 5\,\% uncertainty in the electromagnetic coupling at 100~GeV,
the allowed region of $M_{\rm NC}$ starts already at a few TeV, depending on the compactification scale.

\section{Vacuum birefringence - a remnant effect of high scale noncommutativity}\label{birefringence}
In the last Sect. \ref{powerlaw} we have already seen that a modification of the theory at high energy scales
(there it was the introduction of extra dimensions) can alter the behavior of the noncommutative field theory at
infrared scales. Therefore it makes sense to investigate the consequences of a
modification at a high energy scale $\Lambda \sim M_{\rm{P}}$ even for a 4 dimensional theory.
A simple and natural possibility to model a non-local UV-finite microscopic theory like, e.g., string theory,
is to simply cut off all fluctuations with momenta larger than $\Lambda$
(for a more detailed discussion of this choice see \cite{AJKR2}).

As we will see this cutoff softens the problem of the unwanted mass term for
the photon considerably. Instead of a mass term one has vacuum birefringence
at low momentum scales. If $M_{\rm{NC}}$ is close enough to the cutoff scale
$\Lambda\sim M_{\rm{P}}$ this vacuum birefringence can be pushed beyond the current experimental limits.
Thereby a window for $M_{\rm{NC}}$ opens
where noncommutativity is still allowed.
As experimental and observational sensitivity is likely to improve in the near future this provides an interesting probe for scales $M_{\rm{NC}}$ very close to the Planck scale.

In the following we will restrict ourselves to the case of a pure U(1) noncommutative gauge theory. The discussion of the previous sections shows how this can be generalised to more realistic situations where the photon gets an admixture of a trace-U(1).

Let us now cut off the fluctuations with momenta larger than $\Lambda$ by introducing a factor of $\exp(-\frac{1}{\Lambda^{2}t^{2}})$ in the integral over the Schwinger time $t$. One obtains (s. \cite{Alvarez-Gaume:2003}),
\begin{eqnarray}
\Pi_{\mu\nu}(p)&=& {1 \over
\pi^2}\left(p^2\delta_{\mu\nu}-p_{\mu}p_{\nu}\right) \nonumber \\
&  & \,\,\times \,\,\sum_{j}\alpha_{j}\int_{0}^{1}dx\,
\left[4C(j)-(1-2x)^2 d(j)\right]
\left[K_{0}\left({\sqrt{\Delta_{j}}\over \Lambda}\right)
-K_{0}\left({\sqrt{\Delta_{j}}\over \Lambda_{\rm eff}}\right)\right]
\nonumber\\
&+& {1\over (\pi)^2}\tilde{p}_{\mu}\tilde{p}_{\nu}\,\Lambda_{\rm
eff}^2\sum_{j}\alpha_{j}d(j)
\int_{0}^{1}dx\,\Delta_{j}K_{2}\left({\sqrt{\Delta_{j}}\over
\Lambda_{\rm eff}}\right) \nonumber \\
&+& \delta_{\mu\nu}\left[\mbox{ gauge non-invariant term }\right].
\label{polarization}
\end{eqnarray}
We will neglect the gauge non-invariant terms in the following. They could be treated and eliminated
by using modified Ward-Takahashi identities \cite{Reuter:1993kw,Ellwanger:iz,Freire:2000bq}.

The employed regularisation cuts off the modes $p\gtrsim\Lambda$ in the loop integral in a smooth way. Of course there are lots of different possibilities to do this. Since universality does not hold, different regularisations will in principle lead to different results. However, as long as we leave the qualitative feature ``all momenta $p\gtrsim \Lambda$ are cut off'' holds, we expect that the qualitative results we obtain remain true.

Let us first concentrate on $\Pi_{1}$, i.e. the running gauge coupling.

\begin{figure}
\begin{center}
\scalebox{0.95}[0.95]{
\begin{picture}(190,180)(40,0)
\includegraphics[width=9.5cm]{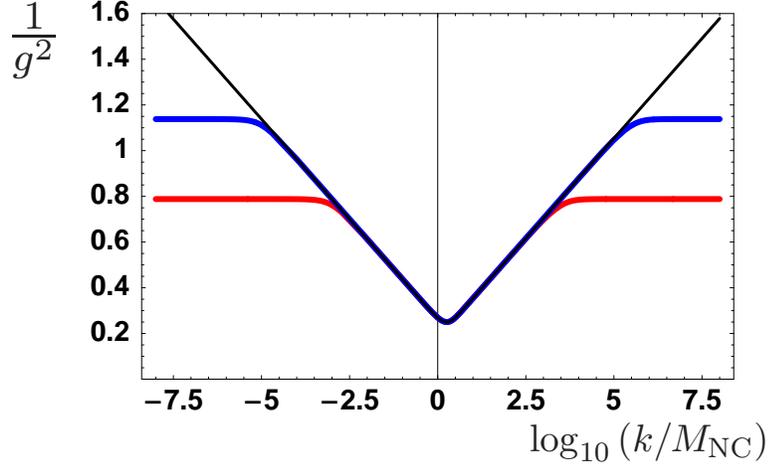}
\Text(-40,-10)[c]{\scalebox{1.4}[1.4]{$\log_{10}\left(k/M_{\rm{NC}}\right)$}}
\Text(-285,150)[c]{\scalebox{2.0}[2.0]{$\frac{1}{g^2}$}}
\end{picture}
}
\end{center}
\caption{Running gauge coupling for a massless supersymmetric pure U(1)
gauge theory. The blue, red and black line are for
$\Lambda=1000\, M_{\rm{NC}}, \,10^5 \,M_{\rm{NC}},\,\infty \,M_{\rm{NC}}$, respectively.
We have fixed the maximal gauge coupling to be $g^{2}_{max}=4$. One can clearly see that for finite values
of the cutoff the running stops at $\sim \Lambda$ in the UV and at $\sim \frac{M^{2}_{\rm{NC}}}{\Lambda}$
in the IR.} \label{gauge}
\end{figure}

In Fig. \ref{gauge} we plot the running gauge coupling for various values of the cutoff $\Lambda$.
As expected the running stops at the UV scale $\Lambda$. In an ordinary commutative theory
we would expect no further changes. Here, however, we observe that the running stops, again, at an
infrared scale $\sim \frac{M^{2}_{\rm{NC}}}{\Lambda}$.
Therefore the running for $k< \frac{M^2_{\rm{NC}}}{\Lambda}$ is essentially the same as that
of a commutative U(1) gauge theory.

Let us now turn to the $\Pi_{2}$ part of the polarisation tensor. It, too, is affected by
the presence of a finite UV cutoff.
For softly broken broken we can easily derive the
following approximate expressions
\begin{eqnarray}
\label{approximate}
\Pi_{2}\!\!&=&\!\!D \Delta M^{2}_{\rm{SUSY}}
\quad\quad\quad\,\,\,\rm{for}\quad \frac{M^{2}_{\rm{NC}}}{\Lambda}\ll k\ll \Delta M_{\rm{SUSY}}
\\\nonumber
\Pi_{2}\!\!&=&\!\! D^{\prime} \Delta M^{2}_{\rm{SUSY}}\tilde{p}^2\Lambda^2
\quad\,\,\, \rm{for}\quad k\ll\frac{M^{2}_{\rm{NC}}}{\Lambda},
\end{eqnarray}
where $D,D^{\prime}$ are known constants.
Following the arguments of Sect. \ref{eom} we can now solve the equations of
motion for the two transverse photon polarisations (noncommutativity matrix as given in Eq. (\ref{thetamatrix})),
\begin{eqnarray}
(\Pi_{1}k^2-\Pi_{2})A^{\mu}_{1}\!\!&=&\!\!0
\\\nonumber
\Pi_{1}k^2 A^{\mu}_{2}\!\!&=&\!\!0.
\end{eqnarray}
Let us now concentrate on the polarisation state $A^{\mu}_{1}$ which is affected by the presence of $\Pi_{2}$.
Inserting the approximate expressions (\ref{approximate}) we can now study the dispersion relation,
\begin{eqnarray}
\label{limit1}
&&\!\!\!\!k^2-D \frac{\Delta M^{2}_{\rm{SUSY}}}{\Pi_{1}}=0,
\quad\quad\quad\quad\quad\quad\quad\,\,\, \rm{for}\quad\frac{M^{2}_{\rm{NC}}}{\Lambda}\ll k\ll \Delta M_{\rm{SUSY}},
\\
\label{limit2}
&&\!\!\!\!k^2+D^{\prime} \frac{\kappa^2}{\Pi_{1}} \frac{\Delta M^{2}_{\rm{SUSY}}\Lambda^{2}}{M^{4}_{\rm{NC}}} (k^3)^2=0,
\quad\quad\quad\!\rm{for}\quad k\ll\frac{M^{2}_{\rm{NC}}}{\Lambda}.
\end{eqnarray}

Eq. (\ref{limit1}) yields the Lorentz symmetry violating mass term of the order of $\Delta M^{2}_{\rm{SUSY}}$
already discussed in detail in Sect. \ref{eom}. Without cutoff, i.e. in the limit $\Lambda\to\infty$
this mass term persists down to $k\to 0$. Thereby excluding any chance that this can be the photon observed
in nature. In presence of the cutoff Eq. (\ref{limit1}) is only applicable
for $k\gg \frac{M^{2}_{\rm{NC}}}{\Lambda}$. Masslessness of the photon is well tested up to at
least $1\,\rm{GeV}$. Using $\Lambda \sim M_{\rm{P}}\sim 10^{18} \,\rm{GeV}$
this gives us a conservative lower bound of $M_{\rm{NC}}>10^{9}\,\rm{GeV}$. Nevertheless,
this opens a rather large window of opportunity compared to the $\Lambda\to\infty$ case where
there was no allowed range of $M_{\rm{NC}}<M_{\rm{P}}$.

For small photon momentum Eq. (\ref{limit2}) applies. To understand (\ref{limit2}) better, let us restore
the light speed $c$ in our equations and use $k^{0}=\omega$ for the frequency of the wave,
\begin{eqnarray}
\label{speed}
\omega^2-c^2(1-\Delta n)^2 (k^3)^2=0,
\end{eqnarray}
with
\begin{eqnarray}
\Delta n
\frac{D}{2} \frac{\kappa^2}{\Pi_{1}} \frac{\Delta M^{2}_{\rm{SUSY}}\Lambda^{2}}{M^{4}_{\rm{NC}}}
\sim 10^{-34}\left(\frac{\Lambda /10^{18}{\rm{GeV}}}{M_{\rm{NC}}}\right)^{4}\ll 1,
\end{eqnarray}
where we have used
$\Delta M^{2}_{\rm{SUSY}}\sim 10^{3}\rm{GeV}$ and $\Pi_{1}\sim 100$.

From Eq. (\ref{speed}) we can see that the photon $A^{\mu}_{1}$ propagates with a
speed $c(1-\Delta n)$. Since the $A^{\mu}_{1}$ photon propagates with $c$ we observe
birefringence, i.e. different polarisations propagate with different speed.

Although $\Delta n$ seems to be quite small we should compare this to the current experimental sensitivity.
In \cite{Kostelecky:2002hh} a study of all possible dimension four Lorentz violating operators in electrodynamics was
conducted and constraints derived.
The most general dimensions four Lagrangian which is gauge and CPT invariant but violates Lorentz symmetry is,
\begin{equation}
\label{general}
{\mathcal{L}}_{\rm{general}}=-\frac{1}{4}F^{\mu\nu}F_{\mu\nu}-\frac{1}{4}(k_{F})_{\mu\nu\alpha\beta}F^{\mu\nu}F^{\alpha\beta}.
\end{equation}
Comparing the propagator derived from (\ref{general}) with Eq. (\ref{poltensor}) we find
\begin{equation}
(k_{F})_{\mu\nu\alpha\beta}=\frac{D}{2}\Delta M^{2}_{\rm{SUSY}}\Lambda^{2}\theta_{\mu\nu}\theta_{\alpha\beta}.
\end{equation}
In \cite{Kostelecky:2002hh} the coefficients of $k_{F}$ have been constrained using various methods. For laboratory measurements their estimate translates to,
\begin{equation}
\Delta n_{\rm{lab}}\lesssim 10^{-10}-10^{-14},
\end{equation}
depending on the pattern of the noncommutativity.
Astrophysical obervations already provide a much tighter bound of
\begin{equation}
\Delta n_{\rm{astro}}\lesssim 10^{-16}
\end{equation}
while the strongest constraints come from observations of objects at cosmological distances
(see also \cite{Kostelecky:2001mb})
\begin{equation}
\Delta n_{\rm{cosmo}}\lesssim 10^{-32}.
\end{equation}

\section{Conclusions}
Noncommutative gauge symmetry
in the Weyl-Moyal approach
leads to two main features which have to be taken into account for sensible
model building. First, there are strong constraints on the dynamics and the field content.
The only allowed gauge groups are U($N$).
In addition, the matter fields are restricted to transform as fundamental, bifundamental and adjoint
representations of the gauge group.
Second, there are the effects of ultraviolet/infrared mixing.
Those lead to asymptotic infrared freedom of
the trace-U(1) subgroup and, if the model does not have unbroken supersymmetry, to Lorentz symmetry violating terms
in the polarisation tensor for this trace-U(1) subgroup.

For a 4 dimensional continuum theory we have demonstrated that, although the trace-U(1) decouples in the limit $k\to 0$, the coupling is not negligibly small
at finite momentum scales $k$, as they appear, for example, in scattering experiments. Therefore,
observations rule out additional unbroken (massless) trace-U(1) subgroups.

Noncommutativity explicitly breaks Lorentz invariance. Therefore an additional \linebreak
Lorentz symmetry violating
structure is allowed in the polarisation tensor. This structure is absent only in
supersymmetric models. If supersymmetry is (softly) broken, this additional structure is present
in the polarisation tensor of the trace-U(1).
It leads to an additional mass $\sim \Delta M^2_{\rm{SUSY}}$ for one of the transverse polarisation states
\cite{Alvarez-Gaume:2003}.
The tight constraints on the photon mass therefore exclude trace-U(1)'s as a candidate for the photon.
It turns out that even a small admixture of a trace part to a traceless part (unaffected by these problems) is fatal.
The only way out seems to be the construction of the photon from a completely traceless generator.
A group theoretic argument shows, that this is impossible whithout having additional
unbroken U(1) subgroups. However, those are already excluded from the arguments given above.

This result severely restricts the possibilities to construct a noncommutative Standard model extension.
If all of the constraints given at the beginning are fulfilled the noncommutativity scale is pushed
to scales far beyond $M_{\rm{P}}$.

In general there is no reason to assume that the simple noncommutative model
used here describes correctly the physics at energies ranging from a few eV up to the Planck mass.
In fact, due to the ultraviolet/infrared mixing, a different ultraviolet embedding of the theory would modify the theory
not only in the ultraviolet, but also in the infrared which can drastically alter these conclusions. E.g., a powerlike decoupling of the trace-U(1) can effectively hide them from observation.
We have demonstrated that in a noncommutative U($N$) gauge theory with compact extra dimensions,
the ultraviolet/infrared mixing effects lead to such a fast power-like decoupling of the
trace-U(1) degrees of freedom. In such a setting the bounds are weakened considerably if the compactification scale is small enough.

As an alternative to extra dimensions we have discussed a modification obtained by simply cutting off all fluctuations
with momenta larger than a cutoff $\Lambda\sim M_{\rm{P}}$.
The presence of an ultraviolet cutoff $\Lambda$ induces an effective infrared scale
$k_{\rm{IR}}\sim \frac{M^{2}_{\rm{NC}}}{\Lambda}$ below which the theory behaves essentially like a commutative gauge theory\footnote{This is in stark contrast to the situation discussed above where the noncommutative gauge theory is assumed to be valid at all scales and no ultraviolet cutoff exists. There $k_{\rm{IR}}=0$ and the
theory shows strong effects of noncommutativity at all scales.}.
In particular, up to threshold effects the running is that of a commutative field theory.
If supersymmetry is broken, we have a Lorentz symmetry violating mass term at scales $k>k_{\rm{IR}}$ in accord with \cite{Jaeckel:2005wt,Alvarez-Gaume:2003}. However, below $k_{\rm{IR}}$ the mass term turns into a modification of the phase velocity of plane wave solutions, leading to birefringence.
Nevertheless, if such a trace-U(1) gauge boson is to be interpreted as (part of) a photon a mass is not acceptable and birefringence must be smaller than the experimental limits.
Using the most stringent limits from cosmological observations
one obtains a rather strong limit of $M_{\rm{NC}}\gtrsim 0.1 M_{\rm{P}}$. If we use the more conservative astrophysical or laboratory limits the same argument yields only
$M_{\rm{NC}}\gtrsim (10^{-7}-10^{-5}) M_{\rm{P}}$. In this setting high precision measurements of the properties of light
are a wonderful tool to test (nearly) Planck scale physics.

\end{document}